
\magnification=\magstep1
\baselineskip= 24 true pt
\vsize=22.5 true cm
\hsize=16.5 true cm
\centerline {\bf Solutions of  quantum Yang-Baxter equation  related to
  $U_q (gl(2))$  algebra }
\centerline {\bf  and  associated  integrable  lattice models }
\vskip  1.25 true cm
\centerline  { B. Basu-Mallick\footnote*{e-mail address:
biru@imsc.ernet.in}}
\centerline  {\it The Institute of Mathematical Sciences }
\centerline {\it  C I T Campus, Taramani, Madras-600113, India }
\vskip 2  true cm
\noindent {\bf Running Title :} Solutions of QYBE related to
                         $U_q (gl(2))$  algebra
\vskip .23 true cm
\noindent {\bf Abstract}

A coloured braid group representation  (CBGR) is constructed with
the help  of some modified universal  ${\cal R}$-matrix,
associated to  $U_q(gl(2))$  quantised  algebra.
Explicit  realisation  of   Faddeev-Reshetikhin-Takhtajan   (FRT)
algebra  is  built  up for this CBGR and new solutions of quantum
Yang-Baxter  equation  are  subsequently found  through
Yang-Baxterisation  of FRT algebra. These   solutions  are
interestingly related to   nonadditive type quantum $R$-matrix and
 have a   nontrivial  $q\rightarrow  1$  limit.  Lax  operators of
several concrete integrable models, which may
 be considered as some `coloured' extensions of lattice nonlinear
Schr${\ddot  {\rm  o}}$dinger  model  and Toda chain, are finally
obtained by taking different reductions of such solutions.

\hfil \eject
\noindent {\bf 1. Introduction }

Quantum  integrable  systems  have  received a lot of interest in
recent years, particularly because  they  represent  a  class  of
nontrivial, exactly solvable models  with many
  possible  physical  applications  [1,2].  A common
feature of each such model  is  the  presence  of  some  mutually
commutating  conserved  quantities including the Hamiltonian, all
of  which  can  be  generated  from the associated Lax  operator.
This Lax operator $L(\lambda ) $ is usually expressed as a matrix having
  noncommuting  operator valued elements, which  are required to
 satisfy the  quantum Yang-Baxter equation (QYBE) given by [2]
$$
R(\lambda , \mu)~ L_1(\lambda)~ L_2(\mu )
{}~ = ~  L_2(\mu )~ L_1(\lambda)~ R(\lambda , \mu),
\eqno (1.1)
$$
where  $~L_1(\lambda  )  =  L(\lambda  )  \otimes  {\bf  1},~ $ $
L_2(\lambda ) = {\bf 1} \otimes L(\lambda ), ~$ and  $~\lambda  ,
{}~\mu  $  are  spectral  parameters. Though $~\lambda , ~\mu $ are
often taken as single component objects, in this article we shall
also investigate on multicomponent spectral parameter dependent  solutions
of  QYBE.  However  such  solutions  may easily be reduced to the
single component case, by considering all but one  components  as
some  dependent  functions  of  the remaining one. The $c$-number
valued elements  of the quantum $R(\lambda , \mu )$-matrix appearing in
(1.1) act like structure constants  and  due to  the  associativity
of  QYBE  they  obey the Yang-Baxter equation (YBE), which may be
written in  the  matrix  form  as
$$
R_{12}(\lambda  ,  \mu  )~
R_{13}(\lambda  ,\gamma  )~R_{23}(\mu  ,\gamma  )~  =~ R_{23}(\mu
,\gamma )~R_{13}(\lambda ,\gamma ) ~R_{12}(\lambda , \mu ), \eqno (1.2)
$$
where we have used the standard notation: $~R_{12} =
R \otimes {\bf 1} ,$ etc. .
 The  QYBE (1.1) plays a crucial role in  determining
 the integrabilily of a system,
 as well as solving it exactly through quantum inverse
scattering method  (QISM)  [2].  Therefore the search for obtaining
  nontrivial solutions  of  QYBE  might  be  considered
as  an important step towards constructing such exactly solvable models.

 Curiously,  quantum  group  related algebras and their different
realisations [3,4] are  found  to  be  intimately  connected  with
certain type of solutions of QYBE and associated  integrable models [5].
  It    is  also   realised    that some basic relations occuring
  in the   well   known
Faddeev-Reshetikhin-Takhtajan (FRT)
  approach  to  quantum  group  [6],  can be used as a
convenient   tool   for   making   such   connection   [7].   These
algebraic    relations  might be written in the form
$$
R^{+}L^{(\pm )}_1 L^{(\pm)}_2 ~=~L^{(\pm)}_2 L^{(\pm)}_1 R^+ ~,~~
R^{+}L^{(+)}_1  L^{(-)}_2 ~=~L^{(-)}_2  L^{(+)}_1 R^+ ~,
\eqno (1.3a,b)
$$
 where  $  ~L_1^{(\pm )}=L^{(\pm )} \otimes {\bf 1},$ $~L_2^{(\pm
)}= {\bf 1} \otimes  L^{(\pm  )}$  and  $L^{(\pm  )}$  are  upper
(lower) triangular matrices with operator valued elements.
   Due   to   the  associativity  of  FRT algebra  (1.3),  the  upper
triangular matrix  $R^+$  satisfies  the  condition
$$  R^+_{12}
{}~R^+_{13}~R^+_{23}~=~R^+_{23}~R^+_{13}~R^+_{12}~,
\eqno  (1.4) $$
which,  in  turn,   leads   to   a   braid   group
representation  (BGR)  for  the matrix ${\hat R}^+ = {\cal P} R^+ $ ( ${\cal
P}$ being the permutation  operator with the property
$~{\cal P} A\otimes B = B
\otimes A {\cal P} $ ): $ {\hat R}_{12}^+ ~{\hat R}_{23}^+ ~{\hat
R}_{12}^+ ~~= $ $~~ {\hat R}_{23}^+ ~ {\hat R}_{12}^+ ~ {\hat R}_{23}^+ . $
     Nevertheless,  for  the  sake  of  convenience  we would call the
$R^+$-matrix  itself,  satisfying  (1.4),  as  the  BGR  in  what
follows.  It may be observed that   the  QYBE (1.1) and FRT algebra (1.3) are
much similar in form, though the latter is void of  any  spectral
parameter. However,
 by  properly  inserting spectral parameters in the FRT relations
through Yang-Baxterisation procedure, it is possible to build  up
for some particular cases an `ancestor' Lax operator $L(\lambda )$
 satisfying  QYBE  [7].  Specific  realisations  of  the operator
elements corresponding to this  $L(\lambda )$ operator,  e.g.  ,
with  physical  bosonic  or  q-bosonic modes, may then lead to
many concrete integrable models  including  some  new  ones  [8].
Noteworthyly, all these integrable models
  are  associated  with  additive type quantum $R$-matrices, which
depend only on the difference of spectral parameters: $ R(\lambda
, \mu ) \equiv R(\lambda - \mu ) $.

 Quite  recently   some  generalisations  of   BGR   (1.4)   have
interestingly  appeared
in   the   literature  [9-11],  which  are  sometimes  called  as
`coloured' braid group representations (CBGRs). These CBGRs  obey
the relation
$$
R_{12}^{ +(\lambda , \mu ) }~ R_{13}^{ +(\lambda ,\gamma ) }~
R_{23}^{ +(\mu ,\gamma ) }~ =~
                            R_{23}^{ +(\mu ,\gamma ) }
     ~R_{13}^{+ (\lambda ,\gamma ) } ~R_{12}^{ +(\lambda , \mu ) },
\eqno (1.5)
$$
where  $~\lambda  ,~  \mu  ,  ~\gamma ~$ are `colour' parameters.
Though  usually  $~{\hat  R}^{+(\lambda  ,\mu  )}  =   {\cal   P}
R^{+(\lambda ,\mu )} $ is defined as the CBGR, we would call the
 $R^{+(\lambda ,\mu )}$-matrix itself as CBGR in analogy with the
previous  standard  case. It may be noticed
that  the  form of eqn. (1.5) is
essentially same with YBE (1.2), if one  interprets  the  `colour
parameters' as `spectral parameters'. Now
  Yang-Baxterisation of FRT algebra related to this CBGR should lead
to new integrable  models, associated with more  general kind of
quantum  $R$-matrix.  This   possibility   was
actually considered in  ref.12 for the case of a particular CBGR given by
$$
R^{ +(\lambda , \mu ) } ~=~ \pmatrix { { q^{1- (\lambda -  \mu
) } } & {} & {} & {} \cr \ { \eqalign { \cr } } & { q^{ \lambda +
\mu  }  } & { (q-q^{-1} ) s^{- (\lambda - \mu ) } } & {} \cr {} &
{0} & { q^{- ( \lambda + \mu ) } } & {} \cr {} & {} & {} & { q^{1
+ (\lambda - \mu ) } }  }  ~,  \eqno  (1.6)
$$
which  might  be
obtained   from   the  fundamental  representation  of  universal
${\cal R}$-matrix corresponding to $U_q(gl(2))$ quantised algebra  [11].
Note that at $\lambda = \mu = 0 $ limit the above CBGR reduces to
the wellknown BGR
$$
   R^+ ~=~ \pmatrix {   {q} & {} & {} & {} \cr {} & {1} &  {q-q^{-1}} & {} \cr
   {} & {0} & {1} & {} \cr {} & {} & {} & {q}  } ~, \eqno (1.7)
$$
 related  to  the $U_q(sl(2))$ algebra. Through Yang-Baxterisaion
of FRT algebra for the case of
 CBGR  (1.6), it is possible to construct as before an `ancestor' Lax
operator as a solution of QYBE. But, contrary to the previous
cases, this Lax operator interestingly  shares  nonadditive  type
quantum $R$-matrix and yields some `coloured'  extensions
        of the integrable systems like lattice sine-Gordon model,
Ablowitz-Ladik model etc. [12].
   In spite of these achievements one obvious drawback in dealing
with  the  CBGR (1.6) is that, its dependence on the colour
parameters $\lambda , ~\mu $ becomes trivial at   the   naive
$q\rightarrow 1 $ limit.  However,
   Lax  operator  of  many  interesting  integrable  systems like
lattice nonlinear Schr${\ddot {\rm o}}$dinger (LNLS) model,  Toda
chain etc. are known to be associated with
   the  rational  limit  (  $q\rightarrow 1 $ ) of the BGR (1.7).
Consequently, one finds it a bit difficult to construct
  `coloured'  generalisation of this type of integrable models
  by starting from the CBGR (1.6). To overcome this problem,
   we  modify  the  previously  obtained co-product and universal
${\cal R}$-matrix related to $U_qgl(2)$ quantum algebra  through  a
scaling  transformation,  as  will  be described in sec.2 of this
article. Fundamental representation of  such  modified  universal
${\cal R}$-matrix then yields a CBGR,
   which  interestingly  has a nontrivial
   colour parameter dependence at $q\rightarrow 1 $ limit
and so is quite suitable for our present  purpose.  In  sec.3  we
build up  explicit  realisations of FRT algebra corresponding to
this new CBGR and in sec.4
   deal  with related Yang-Baxterisation procedure, which leads to new
   solutions of QYBE. The possibility
of generating `coloured' extension  of LNLS or  Toda  chain
type  integrable  models,  associated  with  nonadditive  quantum
$R$-matrix, is also briefly discussed in sec.4.
Sec.5  is  the  concluding
section.
\vskip .5 true cm
\noindent {\bf 2. Construction of a CBGR  from   $U_q(gl(2))$
quasi-triangular Hopf algebra }
\vskip .1 true cm
Before  attempting  to  generate  the CBGR which has a nontrivial
rational limit, we would  review  a  little  about  the
construction   of   CBGR   (1.6)   by  employing  the  techniques  of
quasitriangular  Hopf  algebra.  As  it  is  well  known  for   a
quasitriangular Hopf algebra ${\cal A}$,
there  exists  an invertible universal ${\cal R}$-matrix ( ${\cal
R} \in {\cal A} \otimes {\cal A} $ ) such  that  it  interrelates
comultiplications
 $\Delta ,~{\Delta}'$ through ${\Delta}(a){\cal R}={\cal R} \Delta' (a), $
 where $a \in {\cal A} $  and satisfies the following conditions
 $$
   ( ~id \otimes \Delta ~) { \cal R } ~=~{\cal R }_{13} {\cal R }_{12}~,~~
( ~\Delta \otimes id ~) {\cal R } ~=~{\cal R }_{13} {\cal R }_{23}~,~~
    (~ S \otimes id ~) {\cal R }  ~=~ {\cal R }^{-1}  ~,
$$
 $S$  being the antipode. The above relations also imply that the
${\cal R}$-matrix would satisfy an equation of the form (1.4).

 If  one  considers  now  the case of $U_q(gl(2))$ algebra, apart
from the usual generators  $S_3,~S_{\pm  }$  of  $U_q(sl(2))$,  a
central  element  $\Lambda  $ is included in the picture with the
commutation relations [13,11]
$$
  \eqalign {
[~S_3,S_{\pm }~] ~=~ \pm ~ S_{\pm } , ~~
[~S_+,S_-~] ~=~ { \sin (2\alpha S_3)  \over \sin \alpha } \cr
 \
 [\Lambda , S_\pm ] ~=~  [\Lambda , S_3 ] ~=~0 ~;~~~~~~~  q=e^{i\alpha } . }
\eqno (2.1)
$$
 The  associated comultiplication may be given by
$$
    \eqalign {
    \Delta (S_+ ) ~~&=~~ S_+ \otimes q^{-S_3 } \cdot (qs)^\Lambda
          ~+~   ({s\over q })^\Lambda  \cdot q^{S_3}\otimes S_+ ~,  \cr
    \Delta (S_- ) ~~&=~~ S_- \otimes q^{-S_3 } \cdot (qs)^{- \Lambda }
          ~+~   ({s\over q })^{- \Lambda }  \cdot  q^{S_3} \otimes S_- ~, \cr
  \Delta (S_3 ) ~~=~~ &S_3 \otimes {\bf 1} + {\bf 1} \otimes   S_3 ~, ~~
 \Delta ( \Lambda ) ~~=~~ \Lambda \otimes {\bf 1} + {\bf 1} \otimes \Lambda ~,
}
  \eqno (2.2)
$$
 where  an  additional  parameter  $s$  is  appearing  due to the
symmetry of the algebra. The other Hopf algebraic structures like
co-unit, antipode can be consistently defined and  the  universal
${\cal R}$-matrix may also be constructed as [11]
$$
    {\cal R} = q^{ 2  ( S_3 \otimes S_3  + S_3 \otimes \Lambda -
    \Lambda \otimes S_3  ) } \cdot  \sum_{m=0}^\infty
    \
    {  ( 1-q^{-2} )^m  \over  [m,q^{-2} ] ! } ~ \left ( q^{S_3}
    (qs)^{- \Lambda } S_+ \right )^m  \otimes
    \left (  q^{-S_3} ( {s\over q } )^{ \Lambda } S_- \right )^m ~,
\eqno (2.3)
$$
 where $[m,q] = (1-q^m)/(1-q) $ and $ [m,q]! = [m,q]\cdot [m-1,q]
\cdots 1 $.

 Denoting  now  the  eigenvalue  of  the  Casimir  like  operator
$\Lambda $ by $\lambda $ and  the  corresponding  $n$-dimensional
irreducible  representation of algebra (2.1) as $\Pi^n_\lambda $,
we may obtain the `colour' representation
$(\Pi^n_\lambda   \otimes \Pi^n_\mu  ){\cal R} $,
giving  a finite dimensional CBGR satisfying (1.5). In particular
for the two dimensional representation $\Pi^2_\lambda  $  through
identity operator and Pauli matrices:
 $\Lambda  =  \lambda  {\bf  1},$  $  {\vec S} = {1\over 2} {\vec
\sigma } $ one gets the CBGR (1.6).

 Now  for    constructing  the CBGR which would be useful for our
purpose, we first observe that the $U_q(gl(2))$  quantised
algebra  (2.1)  remains invariant under a scaling transformation of
the generator $\Lambda $  :  $\Lambda  \rightarrow  {1\over  c  }
\Lambda  $,  $c$  being  an arbitrary complex number. However
 the related comultiplication (2.2) interestingly gets
modified due to this transformation.
  Therefore  by using such scaling of generator $\Lambda $, it is
possible to associate a somewhat more general co-multiplication
structure  than (2.2), with the  algebra (2.1). After setting
$t=q^{-c}$ and redefining the parameter $s$ in a
   suitable way, one can cast this modified structure in the form
$$
    \eqalign {
    \Delta (S_+ ) ~~&=~~ S_+ \otimes q^{-S_3 } \cdot (ts)^\Lambda
          ~+~   ({s\over t })^\Lambda  \cdot q^{S_3}\otimes S_+ ~,  \cr
    \Delta (S_- ) ~~&=~~ S_- \otimes q^{-S_3 } \cdot (ts)^{- \Lambda }
          ~+~   ({s\over t })^{- \Lambda }  \cdot  q^{S_3} \otimes S_- ~, \cr
  \Delta (S_3 ) ~~=~~ &S_3 \otimes {\bf 1} + {\bf 1} \otimes   S_3 ~, ~~
 \Delta ( \Lambda ) ~~=~~ \Lambda \otimes {\bf 1} + {\bf 1} \otimes \Lambda ~,
}
  \eqno (2.4)
$$
   where  $q$,  $s$  and $t$  are being treated as three independent
parameters.
   By applying  now  the    scaling  transformation  of generator $\Lambda $
to   the expression (2.3), we may easily obtain the universal
${\cal R}$-matrix related to the  co-product (2.4) as
$$
    {\cal R} ~=~ q^{ 2  S_3 \otimes S_3  } ~
    t^{2 ( S_3 \otimes \Lambda - \Lambda \otimes S_3  ) }  \cdot
     \sum_{m=0}^\infty
    \
    {  ( 1-q^{-2} )^m  \over  [m,q^{-2} ] ! } ~ \left ( q^{S_3}
    (ts)^{- \Lambda } S_+ \right )^m  \otimes
    \left (  q^{-S_3} ( {s\over t } )^{ \Lambda } S_- \right )^m ~.
\eqno (2.5)
$$
Evidently,   for the value  $t=q$  the  above  universal  ${\cal
R}$-matrix would reproduce the previous expression (2.3).
    On  the other hand, at the limit $s=1$ (2.5) curiously coincides
with the case which was obtained very recently [14]
  by  using  Reshetikhin's   procedure of multiparameter deformation.'

After
finding the universal $R$-matrix (2.5), one may  take  as  before
its  two dimensional representation through identity operator and
Pauli matrices. This leads us to the  CBGR
$$
R^{ +(\lambda , \mu ) } ~=~ \pmatrix {
{ q t^{- (\lambda - \mu ) }    } & {} & {} & {} \cr
\
{ \eqalign { \cr  } }
&  { t^{ \lambda + \mu  } }  &  {  (q-q^{-1} ) s^{- (\lambda - \mu ) }   }
& {} \cr
{} & {0} & {  t^{- ( \lambda + \mu ) }  }  &  {} \cr
{} & {} & {} & { q t^{ \lambda - \mu  } }    } ~,
\eqno (2.6)
$$
  which    would  naturally  reduce to (1.6) for  the case  $t =q$. At this
point we may mention about some symmetry  transformation
of  eqn. (1.5)  related  to  the  `particle  conserving'  BGRs  [15,12]. By
applying such a transformation on the
BGR (1.7) it is also possible to obtain the CBGR (2.6), without going into the
associated Hopf algebra structures.

It is worth noticing that
the  colour parameter dependence   of the CBGR (2.6) remain
nontrivial even at the $q\rightarrow 1$ limit.  So  according  to
our  discussion  in  sec.1,  this  newly obtained CBGR should help
us in  constructing  `coloured'
extension   of  integrable  models  related  to  the rational case.
For this purpose  we  shall next  build  up
explicit  realisation  of FRT algebra related to this CBGR.
\vskip .5 true cm
\noindent {\bf 3. Realisation  of FRT  algebra  related  to  the  CBGR }
\vskip .1 true cm
Due to the
presence of extra colour parameters, the more frequently  used  form of FRT
relations  (1.3)  has  to be modified  consistently [6,12]
for the case dealing with a CBGR $R^{+(\lambda ,\mu )} $:
$$
  \eqalign {
R^{+(\lambda ,\mu )}~L^{(\pm )}_1(\lambda )~ L^{(\pm)}_2 (\mu ) ~
&= ~L^{(\pm)}_2 (\mu ) ~L^{(\pm )}_1(\lambda ) ~ R^{+(\lambda ,\mu )}  \cr
R^{+(\lambda ,\mu )}~L^{(+)}_1(\lambda )~ L^{(-)}_2 (\mu ) ~
&= ~L^{(-)}_2 (\mu ) ~L^{(+)}_1(\lambda )~  R^{+(\lambda ,\mu )} ~,  }
\eqno (3.1a,b)
$$
where  $  ~L_1^{(\pm  )}(\lambda ) = L^{(\pm )}(\lambda ) \otimes
{\bf 1},$ $~L_2^{(\pm )}(\lambda )= {\bf 1}  \otimes  L^{(\pm  )}
(\lambda  )$  and  $  L^{(\pm  )}  (\lambda  )$  being upper (lower )
triangular matrices.  By  using  this  coloured  version  of  FRT
algebra,  explicit  realisations of $L^{\pm }(\lambda )$ matrices
were previously built  up  for  the  CBGR  (1.6)  [12].  Now  for
constructing  similar  realisation  for  case of  CBGR (2.6), we
assume the form of corresponding $L^{\pm }(\lambda )$ matrices as
$$
L^{(+)}(\lambda ) = \rho^{-2\lambda }
  \pmatrix {
{ \eqalign { t^\Lambda  \tau_1^+ \cr  }}    &
{  (ts)^{-\lambda } \cdot  s^{\Lambda }    \tau_{21} }
\cr               {0} &  { t^{-\Lambda }  \tau_2^+ }   },~
L^{(-)}(\lambda ) = \rho^{-2\lambda  }
  \pmatrix {
{ \eqalign { t^\Lambda  \tau_1^-  \cr  } }    &  {0}    \cr
{  (ts)^{\lambda  } \cdot s^{\! -\Lambda }  \tau_{12}  } &
{ t^{-\Lambda }  \tau_2^- }     } .
\eqno  (3.2a,b)
$$
Inserting the above form of $L^{\pm }(\lambda )$ matrices as well
as  CBGR  (2.6)  in (3.1), we surprisingly find that the coloured
version of FRT algebra reduces finally to the following algebraic
relations   for   the   operators   $~\tau_i^\pm   ,   ~\tau_{ij}
{}~ (i,j=1,2),~\rho  $  and  $  \Lambda  $,  where  evidently  colour
parameters are absent.
$$
    \eqalign {
\tau_i^{\pm }\tau_{ij} ~=~q^{\pm 1 } \tau_{ij} \tau_i^{\pm }~,~~
\tau_i^{\pm }\tau_{ji} ~=~q^{\mp 1} \tau_{ji} \tau_i^{\pm }~,~
{}~ \rho \tau_{12} ~=~ t ~\tau_{12} \rho ~, ~~
  \rho \tau_{21} ~=~ t^{-1} ~\tau_{21} \rho ~,  \cr
\
\left [ \tau_{12}, \tau_{21} \right ] ~=~ -  (q-q^{-1} )
\left (  \tau_1^+ \tau_2^-  -  \tau_1^- \tau_2^+ \right ) ~,~~~
[\Lambda ,\rho   ] =
[  \Lambda ,\tau_i^\pm ] =  [  \Lambda ,\tau_{ij}  ] = 0 ~, }
\eqno  (3.3)
$$
with  all  generators  $  \rho  ,~\tau_i^\pm  $  commuting  among
themselves. It may be verified that in addition  to  the  Casimir
like  element $\Lambda $, there exists other Casimir operators of
the above algebra as
$$ \eqalign { D_1 = \tau_1^+  \tau_1^-  ~,~~
D_2  = \tau_2^+ \tau_2^- ~,~~ &D_3 = \tau_1^+ \tau_2^+ ~,~~ D_4 =
\rho (\tau_1)^c~,  \cr
D_5 = 2 \cos \alpha ~( \tau_1^+ \tau_2^-  & +
\tau_1^- \tau_2^+ ) - [ \tau_{12}, \tau_{21} ]_+  ~,}
\eqno  (3.4)
$$
 where $~c = - \log_q t. $ It is also interesting to notice that
 the subalgebra of (3.3), containing only the generators ${\vec \tau  }$,
 coincides with the FRT algebra related to the $U_q(sl(2))$  case [7].
Now it is rather easy to check
that the choice of generators $~\rho ,~{\vec \tau } $ in the particular form
$$
\rho = t^{S_3},~~
\tau_1^{\pm } = q^{ \pm S_3 },~~ \tau_2^{\pm } = q^{ \mp S_3 },~~
 \tau_{12} = - (q-q^{-1})S_+ ,~~  \tau_{21} =  (q-q^{-1})S_-
 \eqno  (3.5)
 $$
reduces  algebra  (3.3)  to  the  $U_q(gl(2))$  given  as  (2.1).
Consequently, by substituting (3.5) to (3.2a,b) one gets explicit
forms of
 upper  (lower  )  triangular  matrices  $L^{(\pm  )}(\lambda  )$
expressed through the generators
$~S_\pm  ,~  S_3, ~\Lambda $ of the $U_q(gl(2))$ quantised algebra:
$$
\eqalign { &L^{(+)}(\lambda ) ~~=~~ t^{-2 \lambda S_3 }~
  \pmatrix {
{       t^\Lambda  q^{S_3 }     } &
{ ~~  (q-q^{-1})~ (ts)^{-\lambda } \cdot  s^{ \Lambda }~ S_-   }
\cr  {}&{}  \cr
{0~} &  {~~ t^{-\Lambda } q^{- S_3 } }     }   ,  \cr
                       ~ &~ \cr
&L^{(-)}(\lambda ) ~~=~~ t^{- 2 \lambda S_3  }~
   \pmatrix {
{  ~ t^\Lambda  q^{- S_3 }  ~  }
&  { ~~0   }    \cr  {} & {} \cr
{ - (q-q^{-1})~ (ts)^{\lambda } \cdot  s^{ -\Lambda }~ S_+  ~ } &
{~~ t^{-\Lambda } q^{S_3}   }      }  }
\eqno  (3.6)
$$
Notice  that  if one substitutes $~\lambda = 0 = \Lambda $ in the
above expressions of $L^{(\pm )}(\lambda )$ matrices, they  would reduce
to well known $L^{(\pm  )}$  matrices  associated  with  the  BGR
(1.7).  On  the other hand at the limit $q=t$, they  yield the
$L^{(\pm )}(\lambda )$ matrices [12] related to  the  CBGR  (1.6).
 We  may note further that similar to the FRT algebra (1.3)
its colour counterpart (3.1) also exhibits the symmetry  that  if
$~L^{(\pm  ,1  )}(\lambda )~$ and $~L^{(\pm ,2) }(\lambda )~$ are
two independent solution  of  the  algebra  acting  on  different
quantum  spaces,  then  their  matrix  product  $  \Delta L^{(\pm
)}(\lambda )= ( L^{(\pm , 1)}(\lambda )\cdot L^{(\pm ,2)}(\lambda
) ) $ would also be a solution with the  same  CBGR.  Using  this
important  property  and  the  explicit  forms  (3.6) of $L^{(\pm
)}(\lambda  )$  we  may  easily  derive  the     coproduct
structure  of  the   related quantised  algebra. Curiously one finds that,
though  the  $L^{(\pm  )}(\lambda  )$-matrices   contain   colour
parameter  $\lambda  $ in a rather complicated way, the resultant
coproducts for the generators $~S_\pm ,~ S_3 $ and $  ~\Lambda  $
are  free  from  such  parameters  and in fact coincides with the
modified coproduct (2.4) of $U_q(gl(2))$.

Next we may observe that if $R^{+(\lambda ,\mu )}$ be a CBGR
 satisfying   eqn. (1.5),
then the matrix $R^{-(\lambda ,\mu )}$ given by
$$
  R^{-(\lambda ,\mu )}~=~
  {\cal P}~ \left \{ R^{+(\mu ,\lambda  )} \right  \}^{-1 } ~{\cal P}~,
  \eqno (3.7)
  $$
   (  ${\cal P}$ being the permutation operator with the property
$~{\cal P} A\otimes B = B \otimes A  {\cal  P}  $  )  would  also  be
a solution of the same equation.
Moreover if $R^{+(\lambda
,\mu  )}$  satisfies  the  FRT relations (3.1) for a certain choice of
$L^{(\pm )}(\lambda )$-matrices, then by using (3.7)  it  is  rather
easy   to  verify  that  the CBGR $R^{-(\lambda ,\mu )} $ would
automatically satisfy the following complementary  FRT  relations
for the same  choice of $L^{(\pm )}(\lambda )$-matrices:
$$
  \eqalign {
R^{-(\lambda ,\mu )}~L^{(\pm )}_1(\lambda )~ L^{(\pm)}_2 (\mu ) ~
&= ~L^{(\pm)}_2 (\mu ) ~L^{(\pm )}_1(\lambda )  R^{-(\lambda ,\mu )}  \cr
R^{-(\lambda ,\mu )}~L^{(-)}_1(\lambda )~ L^{(+)}_2 (\mu ) ~
&= ~L^{(+)}_2 (\mu ) ~L^{(-)}_1(\lambda )  R^{-(\lambda ,\mu )} .  }
\eqno (3.8a,b)
$$
By applying now the relation (3.7) to the case of particular CBGR (2.6),
one gets the corresponding lower triangular CBGR as
$$
R^{ -(\lambda , \mu ) } ~=~ \pmatrix {
{ q^{-1}  t^{- (\lambda - \mu ) }    } & {} & {} & {} \cr
\
{ \eqalign { \cr  } }
&  { t^{ \lambda + \mu  } }  &  {0}
& {} \cr
{} & { - (q-q^{-1} ) s^{\lambda - \mu  }   }
& {  t^{- ( \lambda + \mu ) }  }  &  {} \cr
{} & {} & {} & { q^{-1} t^{ \lambda - \mu  } }    } ~.
\eqno (3.9)
$$
So this CBGR  naturally satisfies the complementary FRT relations (3.8) and
the form of related   $L^{(\pm )}(\lambda )$ matrices
 would again be given by (3.6).
\vskip .5 true cm
\noindent {\bf 4. Yang-Baxterisation of coloured FRT algebra
and related integrable models }
\vskip .1 true cm
           As mentioned already, it is possible to construct
           spectral parameter dependent solutions of QYBE (1.1),
through Yang-Baxterisation of FRT algebra related to the standard BGRs [7].
Such solutions may then be employed  to  build up  in a coherent way
Lax operators of a class of quantum integrable systems,
related to trigonometric  quantum $R$-matrix.
Moreover by taking $q\rightarrow 1 $ limit of these solutions,
one can also generate  another class of integrable
systems including well known LNLS model, Toda chain, etc., all
of which interestingly   share rational and additive type quantum $R$-matrix
[8].
The above procedure for constructing integrable models should also
  be applicable to the case of CBGRs, through Yang-Baxterisation
  of corresponding   FRT relations ((3.1),(3.8)).
           An initial attemt was   made  in this direction [12]
  for the case of particular CBGR (1.6),
and consequently some `coloured' extension of lattice sine-Gordon
  model, Ablowitz-Ladik model etc. were obtained. However due
  to the apparent difficulty in extracting  nontrivial
  $q\rightarrow 1$ limit of the
CBGR (1.6), possible `coloured' extension of integrable systems
like LNLS model or Toda chain have not been considered so far.
At present we aim to fill up this gap by  constructing the Yang-Baxterisation
  of FRT algebra related to the modified CBGR  (2.6)
  and then properly switching over
  to the corresponding  $q\rightarrow 1$ limit.
    So,  in parallel to  the approach of  ref.12, at  first we seek
    solutions of QYBE (1.1)
    for the two-component spectral parameter case:
$$
R(\lambda ,  \lambda'; \mu ,\mu' )~ L_1(\lambda ,\lambda')~ L_2(\mu  ,\mu')
{}~ = ~  L_2(\mu ,\mu')~ L_1(\lambda ,\lambda' )~
R(\lambda ,  \lambda' ; \mu ,\mu' )~,
\eqno (4.1)
$$
by substituting in  it  $R(\lambda ,  \lambda'; \mu ,\mu' )$
and  $L_1(\lambda ,\lambda')$ matrices of the form
$$
\eqalignno {
R( \lambda , \lambda' ;  \mu , \mu' )  ~~&=~~
q^{(\lambda' - \mu' ) } R^{+(\lambda , \mu )}  ~-~
         q^{ - ( \lambda' - \mu'  ) } R^{-(\lambda , \mu )} ~, & (4.2)   \cr
L(\lambda ,\lambda' ) ~~&=~~ q^{\lambda' } \cdot  L^{(+)} (\lambda )
  + q^{- \lambda' } \cdot  L^{(-)} (\lambda )~,  & (4.3) }
$$
  where the CBGRs $R^{\pm (\lambda , \mu )}$ and the corresponding
                        upper (lower) triangular matrices
$ L^{(\pm)} (\lambda )$  are given through  the expressions
((2.6), (3.9)) and (3.2a,b) respectively.
Notice  that the  $L$-operator (4.3) is depending  on
two independent spectral parameters $\lambda $ and $\lambda' $, the
first one is related to the colour parameter of FRT algebra, while
the latter one  comes through  Yang-Baxterisation scheme.
To verify that $R$-matrix
(4.2) and $L$-operator (4.3) thus constructed are the solutions of QYBE,
we insert them to (4.1) and match the coefficients of different
powers in spectral parameters $\lambda',~\mu'. $ As a result we obtain a
set of algebraic relations  independent of parameters
$\lambda' ,~\mu'$ and observe that all of these relations, except one,
coincide with the coloured FRT relations ((3.1),(3.8) )
and hence are naturally satisfied by construction.
The only remaining equation is
$$     \eqalign {
   R^{+(\lambda ,\mu )}~L^{(-)}_1(\lambda )~ L^{(+)}_2 (\mu ) ~&-~
R^{-(\lambda ,\mu )}~L^{(+)}_1(\lambda )~ L^{(-)}_2 (\mu )  ~=~  \cr
 &L^{(+)}_2 (\mu ) ~L^{(-)}_1(\lambda )~ R^{+(\lambda ,\mu )} ~-~
L^{(-)}_2 (\mu ) ~L^{(+)}_1(\lambda )~ R^{-(\lambda ,\mu )} .  }
 \eqno (4.4)
$$
By substituting  explicit forms of $ L^{(\pm)} (\lambda )$ (3.2a,b) and
 $R^{\pm (\lambda ,\mu )}$ ((2.6),(3.9)) in the above equation and using the
     operator algebra (3.3),  one can directly verify that this remaining
equation
     would also be satisfied. So we may conclude  that
  $R( \lambda , \lambda' ;  \mu , \mu' )$ and   $L(\lambda ,\lambda' )$
  matrices, given through expressions  (4.2) and (4.3) respectively,
  are indeed a solution of two component spectral parameter dependent
  QYBE (4.1).

However, as remarked earlier, single-component dependent
  solutions of QYBE (1.1) are more frequently used for generating
  concrete integrable models and
  to extract such solutions  at the present  case we may consider the colour
  parameters $\lambda ,\mu $  not as independent ones, but
  as some functions of other spectral parameters
$\lambda' ,\mu' .$   For simplicity one may choose
$~\lambda  ~=~ \theta \lambda' $, $~  \mu  ~=~\theta \mu'  ,$
where $\theta $ is a  constant parameter and may set $ \Lambda = 0 $.
Evidently for this simple choice
the two-component dependent solution (4.2), (4.3)
would reduce to a single-component dependent  one,
which may be explicitly  written as
$$
\eqalign {   &R_1( \lambda' ,  \mu' )  ~=~ \cr
 &\pmatrix {
{  t^{-  \theta (\! \lambda' \!  - \! \mu' \! ) }
a(\! \lambda' \! -  \! \mu'  \! )   }
& { \eqalign {  \cr } } & {} & {} \cr
\
{ \eqalign {  \cr } }  &
{ t^{  \theta ( \! \lambda' \! +  \! \mu' \!)   }
{}~b (\! \lambda' \! -  \! \mu' \! )   } & { q^{\! \lambda' \! -  \! \mu' \! }
  s^{-\theta (\! \lambda' \! -  \! \mu' \! )  }  }    &  {}
\cr
{\eqalign {  \cr } }  & { q^{- (\! \lambda' \! -  \! \mu' \! ) }
s^{\theta (\! \lambda' \! -  \! \mu' \! )  }  }
& {  t^{  - \! \theta (\! \lambda' \! +  \! \mu' \! )    } ~
b ( \! \lambda'  \! -  \! \mu' \! )    } &  {} \cr {} & {} & {} &
{  t^{ \theta \! ( \! \lambda' \! -  \! \mu' \!) }
a ( \! \lambda' \! -  \! \mu'  \!)   }}   }
\eqno (4.5)
$$
where  $ a(\lambda ) = { \sin \alpha ( 1+\lambda ) \over \sin \alpha } ,~$
 $ b(\lambda ) = { \sin \alpha \lambda  \over \sin \alpha } $  and
$$
L_1(\lambda' ) ~~=~~
      \rho^{-2 \theta \lambda'  }  ~
      \pmatrix {
{  q^{\lambda' } \cdot \tau_1^+ + q^{- \lambda' } \cdot \tau_1^-     }
&
{ q^{ \lambda' } (st)^{- \theta \lambda' }   \cdot \tau_{21}  }
\cr
{   q^{ - \lambda' } (st)^{ \theta \lambda' }  \cdot \tau_{12}  }
&  {  q^{\lambda' } \cdot \tau_2^+ + q^{- \lambda' }\cdot \tau_2^-     }  }.
\eqno (4.6)
$$
Now for  constructing `coloured'
extension of integrable models related to the rational case, we
intend to study the $q\rightarrow 1$ limit of the above solution.
Evidently, at this limit the matrix
 $R_1( \lambda' ,  \mu' )$  (4.5) yields
$$
\eqalign {
&R_2(\lambda',\mu' ) ~=~ \cr
&\pmatrix {
{ (1 \!+\!\lambda'\! -\!\mu' )~ t^{ - \theta  (  \lambda'  -   \mu' ) }   }
& {} & {} & { } \cr
{ } & { (\lambda'\! - \!\mu' ) ~t^{  \theta  (  \lambda'  +   \mu' ) }   } &
{  s^{- \theta ( \lambda'  -   \mu'  )  }  } & {} \cr
{} & {  s^{ \theta ( \lambda'  -   \mu'  )  }  } &
 { (\lambda'\! - \!\mu' ) ~t^{ - \theta  (  \lambda'  +   \mu' ) }   }  & {}
\cr
 {} & {} & {} &
 { (1 \!+ \!\lambda' \!-  \!\mu' ) ~
 t^{  \theta  (  \lambda'  -   \mu' ) }   }  } }
\eqno (4.7)
 $$
Notice that  though the above
$R_2(\lambda',\mu' )$   is  not rational in form, for the case
  $\theta = 0 $,  it
 interestingly reduces to the well known  additive, rational $R$-matrix
  related to the LNLS model or Toda chain.
So one may naturally expect that this $R_2(\lambda',\mu' )$ can be linked with
 some `coloured' extension  of such integrable models.
However, the   extraction of $q\rightarrow 1$ limit for the  case of
   $L_1(\lambda' )$ operator (4.6) turns out to be  a little difficult,
      since at this limit the associated algebra  (3.3) would
     become almost trivial and cannot be used to produce any physically
     interesting integrable model. To bypass  this difficulty
     we consider a new set of operator  elements   ${\vec K}$,
 which can be defined by inverting  the relations
 $$
     \tau_1^\pm = { K_1 \sin  \alpha \pm K_2  \over 2 \sin \alpha  },~~
     \tau_2^\pm = { K_3 \sin  \alpha \pm K_4  \over 2 \sin \alpha  },~~
     \tau_{12} = K_+ , ~~\tau_{21} = K_- ~.
     \eqno (4.8)
 $$
     Now  by using (4.8),  one  can
    rewrite the algebra (3.3) in terms of the new elements  ${\vec K}$
     and  $\rho $. Interestingly, the $q\rightarrow 1 $ limit of such algebra
    turns out  be a nontrivial one and may be given by
$$
 \eqalign {
     & [ K_1 , K_{\pm } ] = \pm K_\pm K_2 ,~
  [ K_3,  K_{\pm } ] = \mp K_\pm K_4 ,~[K_+ ,K_- ] = (K_1 K_4 - K_2 K_3 ), \cr
 & [\rho , K_1 ] = [ \rho , K_3 ] =
  [ K_1 ,K_3 ] = 0 ,~ \rho K_+ = t K_+ \rho , ~\rho K_- = t^{-1} K_- \rho , }
  \eqno (4.9)
$$
 where  $K_2, ~K_4 ~$ are commuting with all elements of the algebra.
  Next  we also express $L_1(\lambda' )$  operator  (4.6)
  through the new elements appearing in (4.8)
  and then take  the required $q\rightarrow 1 $ limit,  which yields
$$
L_2(\lambda' ) ~~=~~ \rho^{-2 \theta \lambda'  }  ~
      \pmatrix {     { K_1 + i \lambda' K_2 } &
{  (st)^{ - \theta \lambda' }   K_-  }
\cr
{ (st)^{\theta \lambda' }  K_+  } &  {  K_3 + i \lambda' K_4 }   }.
\eqno (4.10)
$$
  Evidently $R_2(\lambda' , \mu' )$ (4.7) and  $ L_2(\lambda' )$ (4.10)
  form a  single-component spectral parameter dependent solution of QYBE (1.1),
  if the corresponding  operator elements satisfy the algebra (4.9).

  Now for constructing some  concrete integrable models
  we may observe that, the subalgebra  of (4.9) involving only
  the elements ${\vec K}$ would reduce to standard $sl(2)$ algebra:
  $ [ K_1 , K_{\pm } ] = \pm K_\pm  , ~$
  $ [K_+ ,K_- ] = 2 K_1  ,$  for the particular case
  $ K_2 = K_4 = 1 $ and  $K_1 = - K_3 $. So in analogy with
  the Holstein-Primakoff  transformation  of  $sl(2)$ generators, one can
   realise the  algebra (4.9) for the above mentioned reduction  as
  $$
  K_1 =  s - N  , ~~\rho = t^{ -N} ,
  ~~ K_+ = (2s - N )^{1\over 2 } a ,  ~~ K_- = a^\dagger  (2s - N )^{1\over 2
},
  \eqno (4.11)
  $$
  where   $s$ is a constant  parameter, $N = a^\dagger a $
  and  $a,~a^\dagger $ satisfy bosonic commutation relations:
  $[a, a^\dagger ] =1 $.  By substituting now  the above realisation
  in $L_2(\lambda' )$ (4.10),   one will readily obtain a concrete Lax operator
  expressed through the physical bosonic modes $a,~a^\dagger $.
  This  Lax operator
  would automatically satisfy the QYBE (1.1), if  the corresponding quantum
  $R$-matrix is taken  as (4.7).  Moreover, for the special case
 $\theta = 0$ it will  surprisingly
    reduce to the Lax operator of quantum integrable  LNLS model [16]
after some trivial transformation. So such Lax operator, as obtained
through the  Holstein-Primakoff type
realisation (4.11), is expected to generate some `coloured'
extension of familiar  LNLS model.

     Next we may construct another simple
     realisation of   algebra (4.9) in the form
$$
K_1 = p ,~~ K_2 = i ,~~ K_3 = K_4 = 0 ,~~ \rho = e^{\beta p },
{}~~K_{\pm } = e^{\mp u },  \eqno (4.12)
$$
where  $\beta = -i ~ \ln t $ and
the canonical operators  $u,~p$ satisfy $~[u,p] = i $. Substitution of
this realisation in (4.10) would again generate a Lax operator, which may
be written explicitly  as
$$
L(\lambda' ) = e^{k \lambda' p } \pmatrix {
{p -\lambda' }  &  { (st)^{-\theta \lambda' } e^u } \cr
 { (st)^{\theta \lambda' } e^{-u} } &  {0}  } ~,
 \eqno (4.13)
 $$
 where $k =-2\theta \beta $.
Notice that at the limit $\theta = 0 $, (4.13)
reduces to the Lax operator of well known Toda chain [17].  So the Lax operator
(4.13), associated with the nonadditive $R$-matrix (4.7),  should finally
 lead to  some `coloured' extension  of  this Toda chain.
We may  hope that by considering  similar other   realisations of algebra
(4.9),
it would be possible  to produce  various quantum integrable
models, through their representative Lax operators.
\vskip .5 true cm
\noindent {\bf Conclusion }
\vskip .1 true cm
  The  FRT  algebra  related  to  quantum  group  can  be used to
generate solutions of QYBE through Yang-Baxterisation  procedure.
However  most  of  the  earlier  attempts  in this direction were
restricted to the case of  standard  BGRs,  which  gave  rise  to
additive type quantum $R$-matrices depending on the difference of
spectral parameters.
        With the aim of generating more general solutions of QYBE
and  related  integrable  models,  in  this  article we apply the
Yang-Baxterisation procedure
              to  the  case of a CBGR. For this purpose, we first
modify  the  known  universal  ${\cal   R}$-matrix   related   to
$U_q(gl(2))$ quantised algebra through a scaling transformation
                   of corresponding central element and construct
a CBGR from the fundamental representation
      of  that  universal  ${\cal  R}$-matrix. Interestingly, the
colour parameter dependence of the  CBGR  obtained  in  this  way
remain nontrivial even at $q\rightarrow 1 $ limit.
We  are  able  to  find  also  explicit  form  of elements of the
coloured FRT algebra like $L^{(\pm)}(\lambda ) $, related to this
CBGR. Though these upper or lower triangular matrices  manifestly
depend  on  the  colour parameter, the underlying quantum algebra
and the associated coproduct reproduced by them are found  to  be
standard ones, devoid of colour parameters.

For   obtaining   new   solutions   of   QYBE,   we  subsequently
Yang-Baxterise  the  FRT  algebra  related  to  the  CBGR.   Such
solutions are found to be associated with non-additive type
     quantum   $R$-matrix,   which   curiously  depends  on  both
difference and summation of  spectral  parameters.  Moreover  the
$q\rightarrow  1 $ limit of these solutions turns out to be quite
interesting and leads to several concrete  Lax  operators,  which
would probably generate
       some  `coloured'  extension  of LNLS model and Toda chain.
The problem of constructing Hamiltonian  of  these  models  along
with their energy spectrum through QISM have not been carried out
by us, and should be studied separately.
 We   may   also   hope   that   by   applying  similar  type  of
Yang-Baxterisation procedure to the case of other CBGRs, it would
be possible to find out a rich class of  solutions  of  QYBE  and
integrable models associated with nonadditive quantum $R$-matrix.
\hfil  \break
\vfil \eject
\noindent {\bf References }
\vskip .2 true cm
\item {[1]} Baxter R J 1982 {\it Exactly Solved Models in
Statistical Mechanics } ( London : Academic Press )
\item  {[2]}
Faddeev  L  D  1980 {\it Sov. Sc. Rev. } {\bf C1 } 107 ;
\item {}
Sklyanin E K, Takhtajan L A and Faddeev L D 1979 {\it Teor. Math.
Fiz. } {\bf 40} 194 ;
\item {} Thacker H B 1982 in  {\it  Lecture
notes  in Physics ,} eds. Hietarinta J et al. ( Berlin : Springer
Verlag ) Vol.{\bf 151 } 1
\item {[3]}  Drinfeld  V  G  1986  {\it
Quantum  groups  }  (  ICM Proceedings, Berkeley ) 798 ; \item {}
Jimbo M 1985 {\it Lett. Math. Phys. } {\bf 10 }  63
\item  {[4]}
Macfarlane  A  J  1989 {\it J. Phys. } {\bf A22 } 4581 ;
\item {}
Biedenharn L C 1989 {\it J. Phys. } {\bf A22 } L873
\item  {[5]}
Sklyanin E K 1982 {\it Funct. Anal. Appl. }{\bf 16 } 27;
\item {}
Tarasov  V O, Takhtajan L A and Faddeev L D 1983 {\it Teor. Math.
Fiz. } {\bf 57} 163
\item  {}  Faddeev  L  D  1990  in  {\it  New
problems,  Methods  and Techniques in Quantum Field Theory } eds.
Rasetti  M  G  (  Singapore:  World  Scientific  )  Advances   in
Statistical Mechanics Vol.{\bf 6} 7
\item {[6]}
 Reshetikhin N Yu,  Takhtajan L A  and  Faddeev L D 1989 {\it Algebra
and analysis }  {\bf 1}    178
\item {[7]}
Bazhanov V V and Stroganov Yu G 1990
in  {\it  Yang-Baxter equation  in integrable systems, } M. Jimbo (ed.),
( Singapore: World Scientific )
Advanced series in mathematical physics, Vol.{\bf 10 } 673 ;
\item {} Basu-Mallick  B and  Kundu A  1992  {\it J. Phys. } {\bf A25 } 4147
\item {[8]}
 Kundu A and  Basu-Mallick B 1992 {\it Mod. Phys. Lett. } {\bf A7}    61 ;
\item {}  Basu-Mallick B and  Kundu A 1992  {\it Phys. Lett.  } {\bf B287}  149
;
\item {} Kundu A and  Basu-Mallick B 1993 {\it J. Math. Phys. } {\bf 34 } 1052
 \item {[9]}
   Akutsu Y and  Deguchi T 1991 {\it Phys. Rev. Lett. } {\bf 67}  777 ;
\item {}  Deguchi T and  Akutsu Y 1991 {\it J.Phys. Soc. Jpn. } {\bf  60 } 2559
\item {[10]}
Ge M L, Sun  C P and  Xue K 1992 {\it Int. Jour.  Mod. Phys. }
{\bf A7 }  6609;
\item {} Ge  M L and Xue K 1993 {\it J. Phys. } {\bf A26 } 281
\item {[11]}
 Burdik C and  Hellinger P 1992 {\it J. Phys. } {\bf  A25 }  L1023
\item {[12]}   Kundu A and  Basu-Mallick B 1993
{\it Coloured FRT algebra and its Yang-Baxterisation leading to
integrable models with nonadditive $R$-matrix } Saha Inst. Preprint,
SINP-TNP/93-05
\item {[13]}   Schirrmacher A,  Wess J and  Zumino B 1991 {\it Z Phys. }
{\bf C49 }  317
\item {[14]} Chakrabarti R and Jagannathan R 1993 {\it On } $U_{p,q}(gl(2))$
{\it and a } $(p,q)-${\it Virasoro algebra}, IMSc Preprint, imsc-93/41
\item {[15]}  Kundu A and  Basu-Mallick B 1992 {\it J. Phys. } {\bf A25 }  6307
\item {[16]}
 Izergin A G and Korepin V E 1982  {\it Nucl. Phys. } {\bf B205  [FS 5]  } 401
 \item {[17]}
 Kulish P and Sklyanin E K 1982 in
 {\it Lecture  notes in Physics ,} eds. Hietarinta
J et al. ( Berlin : Springer Verlag ) Vol.{\bf 151 } 61

\end